\documentclass[aps,prl,showpacs,twocolumn,superscriptaddress,preprintnumbers,nobibnotes]{revtex4-1}

\RequirePackage{subfigure}
\usepackage{lipsum}
\usepackage{graphicx}
\usepackage{color}
\usepackage{enumerate}
\usepackage{amsmath}
\usepackage{bm}
\usepackage{diagbox}
\usepackage{amssymb}
\usepackage{stmaryrd}
\usepackage{multirow}
\usepackage{bbm}
\usepackage[normalem]{ulem}
\usepackage{float}
\usepackage{longtable,booktabs}
\usepackage[dvipsnames,table,xcdraw]{xcolor}
\usepackage[normalem]{ulem}
\usepackage{physics}
\usepackage{url}
\usepackage{soul} 
\usepackage[colorinlistoftodos]{todonotes}
\usepackage{verbatim}
\usepackage{graphicx}
\usepackage{subfigure}
\graphicspath{{figures/}}
\usepackage{appendix}
\definecolor{darkblue}{rgb}{0.1,0.2,0.6} \definecolor{darkred}{rgb}{0.8,0.1,0.2}
\usepackage[colorlinks,citecolor=darkblue,linkcolor=darkred,urlcolor=darkblue]{hyperref}
\usepackage[all]{hypcap} 
\usepackage[draft]{pdfcomment}
\usepackage{mathtools}

\newcommand{\cD}{\mathcal{D}}

\newcommand{\cN}{\mathcal{N}}

\newcommand{\cL}{\mathcal{L}}

\newcommand{\bE}{\mathbb{E}}

\newcommand{\bR}{\mathbb{R}}
\newcommand{\bC}{\mathbb{C}}
\newcommand{\psit}{\psi_{\theta}}
\newcommand{\INNQS}{$\infty$-NNQS }

\newcommand{\be}{\begin{equation}}
\newcommand{\ee}{\end{equation}}
\newcommand{\bea}{\begin{eqnarray}}
\newcommand{\eea}{\end{eqnarray}}

\usepackage{graphicx}
\usepackage{dcolumn}
\usepackage{bm}

\begin{document}

\title{Infinite Neural Network Quantum States: Entanglement and Training Dynamics}

\author{Di Luo}
\thanks{Corresponding author: diluo@mit.edu}
\affiliation{The NSF AI Institute for Artificial Intelligence and Fundamental Interactions\vspace{.1cm}}
\affiliation{Department of Physics, University of Illinois at Urbana-Champaign, IL 61801, USA\vspace{.1cm}}
\affiliation{Center for Theoretical Physics, Massachusetts Institute of Technology, Cambridge, MA 02139, USA\vspace{.1cm}}
\affiliation{Department of Physics, Harvard, Cambridge, MA 02139, USA\vspace{.1cm}}
\author{James Halverson}
\affiliation{The NSF AI Institute for Artificial Intelligence and Fundamental Interactions\vspace{.1cm}}
\affiliation{Department of Physics, Northeastern University, Boston, MA 02115}

\begin{abstract}
 We study infinite limits of neural network quantum states ($\infty$-NNQS), which exhibit representation power through ensemble statistics, and also tractable gradient descent dynamics. Ensemble averages of entanglement entropies are expressed in terms of neural network correlators, and architectures that exhibit volume-law entanglement are presented. The analytic calculations of entanglement entropy bound are tractable because the ensemble statistics are simplified in the Gaussian process limit.
 A general framework is developed for studying the gradient descent dynamics of neural network quantum states (NNQS), using a quantum state neural tangent kernel (QS-NTK). For \INNQS the training dynamics is simplified, since the QS-NTK becomes deterministic and constant. An analytic solution is derived for quantum state supervised learning, which allows an \INNQS to recover any target wavefunction. Numerical experiments on finite and infinite NNQS in the transverse field Ising model and Fermi Hubbard model demonstrate excellent agreement with theory. \INNQS opens up new opportunities for studying  entanglement and training dynamics in other  physics applications, such as in finding ground states.
\end{abstract}

\maketitle

\textbf{Introduction.}--- Quantum states are fundamental objects in quantum mechanics. Generically, the dimensionality of a quantum state grows exponentially with the system size, which provides one fundamental challenge for classical simulations of quantum many-body physics. This is the so-called curse of dimensionality, which also regularly arises in machine learning (ML), where a judicious choice of neural network architecture and optimization method can help address the problem.

Inspired by progress in machine learning, neural networks have have been proposed ~\cite{Carleo602} as a useful way to represent quantum wavefunctions, an idea known as a neural network quantum state (NNQS). The goal is to find a compact neural network representation of the high dimensional quantum state, which is possible because the neural network is a universal function approximator \cite{Cybenko1989,HORNIK1991251}; furthermore, they also give exact representations of certain quantum states ~\cite{Gao2017,topo_wf,Levine_2019,sharir2021neural,luo2020gauge,luo2021gauge,dongling, huang_chiral}, demonstrating their representation power. Recent research has demonstrated that NNQS can achieve state-of-the-art results for computing ground states and the real time dynamics properties of closed and open quantum systems across a variety of domains, including condensed matter physics, high energy physics, and quantum information science~\cite{Han_2020,Choo_2019,rnn_wavefunction,Luo_2019,paulinet,PhysRevResearch.2.033429,quantum_circuit,gutierrez2020real,Glasser_2018,Vieijra_2020,Nomura_2017,Schmitt_2020,Stokes_2020,Vicentini_2019,Torlai2018,PhysRevE.101.023304,PhysRevLett.126.032001,PhysRevB.99.214306,PhysRevLett.122.250502,PhysRevLett.122.250501,luo2020gauge,luo2021gauge,qaoa2021,wang2021spacetime,py2021,nuclear2021}. 
Despite this progress, there is ample room for an improved understanding of the representation power and training dynamics of NNQS.

The neural tangent kernel (NTK) \cite{NTK} has recently emerged as a theoretical tool for understanding the gradient descent dynamics of large neural networks. NTK theory utilizes architectures with a discrete hyperparameter $N$, such as the width of a fully-connected network. In general, gradient descent updates to the network are controlled by a parameter-dependent NTK, but in the infinite-$N$ limit the network evolves as a linear model, with dynamics governed in a ordinary differential equation by a deterministic constant NTK \cite{NTK,lee2019wide,roberts2021principles}. This ODE becomes linear and analytically solvable for a mean-squared-error loss (See the Supplementary Material for a review of the NTK). Similarly, in the infinite-$N$ limit, networks are often drawn from Gaussian processes \cite{neal,williams,lee,Matthews2018GaussianPB}, in which case they may be trained with Bayesian inference via another deterministic constant kernel, the neural network Gaussian Process (NNGP) kernel \cite{neal}.

In this work we study infinite neural network quantum states ($\infty$-NNQS), which exhibit both representation power through ensemble statistics and also tractable training dynamics. Specifically,  we relate ensemble averages of entanglement entropy bound to neural network correlation functions. For appropriate $\infty$-NNQS, the ensemble statistics are Gaussian and the correlators are exactly computable. Architectures are presented that approach Gaussian i.i.d. wavefunctions with volume-law entanglement. 
Furthermore, we develop a general framework for the gradient descent dynamics of NNQS, using a quantum state neural tangent kernel (QS-NTK). Our framework is general and may be applied to various learning setup, such as ground state optimization, quantum state tomography and quantum state supervised learning. In appropriate infinite limits, gradient descent of the \INNQS is governed by a constant deterministic QS-NTK. In the case of quantum state supervised learning, we prove that an \INNQS trained with a positive-definite QS-NTK can recover any target wavefunction. We experimentally demonstrate that the QS-NTK can predict the training dynamics of ensembles of finite width NNQS.

\textbf{Infinite Neural Network Quantum States.}--- 
Consider a quantum state $|\psi\rangle$ represented by a 
neural network with continuous learnable parameters $\theta$ and
a discrete hyperparameter $N$. The wavefunction is
$\psi_{\theta,N}: D \to \bC$, 
where the domain $D$ is problem-dependent. The subscripts $\theta,N$ will often be implicit.  

An infinite neural network quantum state ($\infty$-NNQS) is a neural network representation in the $N\to \infty$ limit.
There are many such limits, according to the identification of a candidate $N$ in a given 
network architecture,
We study cases where this limit is useful either for understanding the entanglement of an ensemble of wavefunctions, via increased control over their statistics, or their gradient descent dynamics. 
For instance, in many architectures the $N\to \infty$ limit is also one in which the network is drawn from a Gaussian process (GP), where, e.g., $N$ is the width a
of a fully-connected network \cite{neal,williams,lee,Matthews2018GaussianPB}  or the number of channels in a CNN \cite{Novak2018BayesianCN,GarrigaAlonso2019DeepCN}. The existence of such NNGP limits is quite general \cite{yangTPorig,yangTP1,yangTP2}, and allows for training with Bayesian inference \cite{neal, lee}.

\textbf{Quantum State NNGP and Entanglement.}--- NNQS exhibit unique and interesting entanglement properties \cite{dongling,Levine_2019,rbm_ee,area_law_rbm}. The statistical control offered by this NNGP correspondence allows us to study the entanglement entropy properties of the ensembles of $\infty$-NNQS. Consider an ensemble of normalized NNQS $\{\ket{\psi_{\theta}}\}$. 
We split the input domain $D$ into a subregion $A$ and its complement $B$ as $D = A \cup B$, which makes the wavefunction arguments consisted of two variables $x_A$ and $x_B$ from subregions $A$ and $B$.

Denote the ensemble average of the $n$-th R\'enyi entanglement entropy as $\langle S_n \rangle \equiv \mathbb{E}_{\theta} S_n$, where $S_n = \frac{1}{1-n} \text{log} \text{Tr} \rho_{\theta A}^{n}$ is the $n$-th R\'enyi entropy of the ensemble over a sub-region $A$. According to Jensen's inequality, $\langle S_n \rangle \geq \frac{1}{1-n} \text{log} \mathbb{E}_{\theta} \text{Tr} \rho_{\theta A}^{n}$ for $n > 1$. It provides a lower bound for entanglement entropy which can be computed from $\mathbb{E}_{\theta} \text{Tr}[\rho_{\theta A}^{n}]$ using the replica-trick~\cite{vmc_ee,autoregressive_ee}:
\begin{align}
    \mathbb{E}_{\theta} \text{Tr}[\rho_{\theta A}^{n}] 
    &= \sum_{x^k_A, x^k_B, k} \mathbb{E}_{\theta} [ \prod_{k=1}^{n} \psi_{\theta}(x^{k,k}_{AB}) \psi_{\theta}^{*}(x^{k+1,k}_{AB}) ] \\ 
    &=\sum_{x^k_A, x^k_B, k} G^{(2n)}(x_{AB}^{1,1},x_{AB}^{2,1},\dots, x_{AB}^{n,n},x_{AB}^{1,n}),
\label{eq:renyi}
\end{align}
where $G^{(2n)}$ are the NNQS correlation functions, defined implicitly, and $x_{AB}^{i,j}:=(x_A^i,x_B^j)$ (here we have the convention $x_{A/B}^{n+1}\equiv x_{A/B}^1$). The sum $\sum_{x^k_A, x^k_B, k}$ is over all $k$ and possible $x^k_A$ and $x^k_B$. This provides a means for analyzing the different entanglement entropies. The entanglement entropy bound is particularly tractable for $\infty$-NNQS, since in the GP limit the correlation functions are determined in terms of the two-point function (GP kernel) via Wick's theorem. See the Supplementary Materials for more details.

Consider $\psi(x)=\psi_1(x)+i \psi_2(x)$, where both $\psi_1(x)$ and $\psi_2(x)$ are drawn from any NN architecture. For example, we analyze the Cos-net ~\cite{halverson2021building} NNQS, where $\psi_1(x)$ and $\psi_2(x)$ come from the following function form:

\begin{equation}
    f(x) = \sum_{i=1}^N a_i \sum_{j=1}^d cos(w_{ij} x_j + b_j)
    \label{eq:cosnet}
\end{equation}
where $d$ is the input dimension, $N$ is the number of hidden dimension, $a_i \sim \mathcal{N}(0,\frac{\sigma_a^2}{N}), w_{ij} \sim \mathcal{N}(0,\frac{\sigma_w^2}{d}),b_j \sim \mathcal{U}[-\pi,\pi])$. It has been shown that in the infinite $N$ limit, $f(x)$ gives rise to the following 2-pt function~\cite{halverson2021building}
\begin{equation}
   \mathbb{E}(f(x),f(y))= G^{(2)}(x,y) = \frac{\sigma_a^2}{2} \, e^{-\frac{\sigma_w^2}{2d}(x-y)^2},
\end{equation}

\begin{figure}[t]
\centering
        \includegraphics[width=1\linewidth]{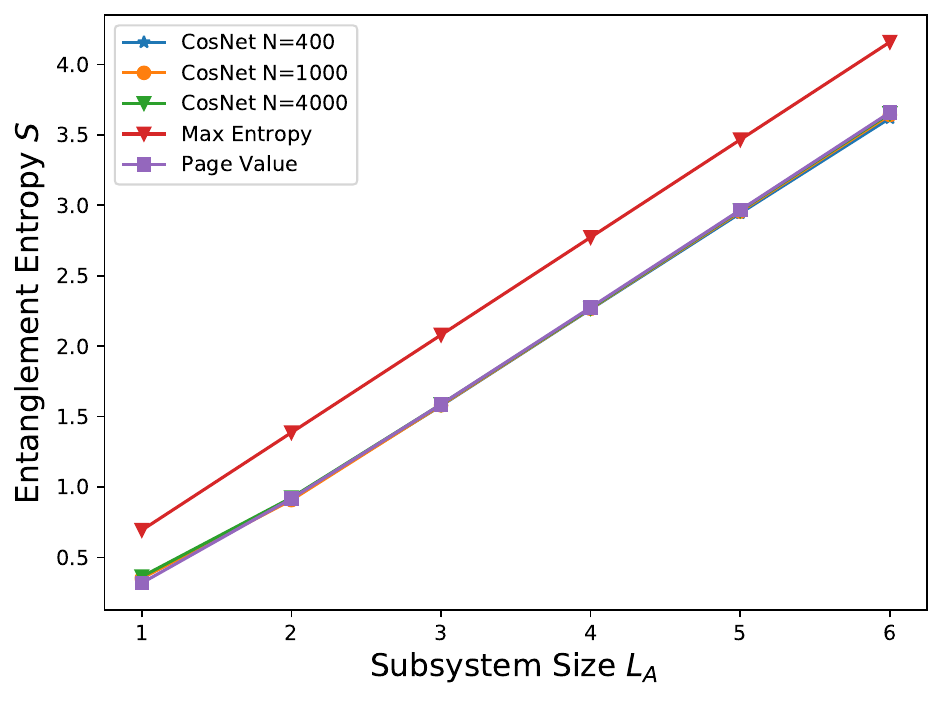}
        \caption{\label{fig:entropy} Von Neumann entanglement entropy of CosNet with width=400,1000,4000 average over 100 ensembles. Max indicates the maximum entropy of the subsystem and Page indicates the page value entropy of the subsystem. }
\end{figure}

 By tuning $\sigma_w \to \infty$, it yields a zero-mean Gaussian process so that $\psi_{1}(x)$ and $\psi_2{(x)}$  are both drawn from i.i.d Gaussian for different values of $x$. After normalization, such an ensemble of wavefunctions is known to reach the Page value of entanglement entropy and  exhibits a volume law entanglement behavior~\cite{Dan_page,entropy}. We compare the Von Neumann entanglement entropy of CosNet with $N=400,1000,4000$ with respect to the Page Value entropy subsystem scaling in Fig.~\ref{fig:entropy}, which demonstrates nice consistency between our theory and simulations. More details on the simulations can be found in the Supplementary Materials. 

 More generally, neural networks provide a means for defining ensembles of wavefunctions with entanglement entropy ensemble average bound expressed in terms of NN correlators even away from the GP limit. This provides a new mechanism for engineering ensembles of wavefunctions whose typical states could have interesting entanglement properties. In general,  finite-$N$ effects introduce non-Gaussianities into the ensemble ~\cite{Yaida:2019sjo,nn_qft} that correct the entanglement entropies. For instance, Gauss-net~\cite{nn_qft} and Cos-net yield dual GPs as $N \to \infty$ ~\cite{jh_toappear}, but have different statistics and even symmetries ~\cite{maiti2021symmetry}  at finite-$N$. It opens up the possibility of entanglement engineering of NNQS and provides a framework for studying entanglement structure of NNQS.

\textbf{Quantum State Neural Tangent Kernel.}--- \INNQS also have interesting gradient descent properties.

We begin with a study of gradient descent for general NNQS. The dynamics of the network are governed by the parameter update $\dot \theta_i = -\nabla_{\theta_i} L = - \sum_{x'\in B} \nabla_{\theta_i} \mathcal{L}(x')$, where we have expressed the update in terms of a total loss $L$ and also a pointwise loss $\mathcal{L}$, summed over a 
batch $B$. Applying the chain rule,
\be
\frac{d\theta_i}{d\tau} = -\eta \sum_{x' \in B} \left[ \frac{\partial \psi(x')}{\partial \theta_i}\frac{\partial \cL}{\partial \psi(x')} + \frac{\partial \psi^*(x')}{\partial \theta_i}\frac{\partial \cL}{\partial \psi^*(x')} \right],
\ee 
where $x'$ is data from $B$ and the loss derivatives are also evaluated on the batch; the structure of $B$ will be further specified in examples, including any labels associated to $x'$. The associated wavefunction update is
\begin{align}
\frac{d\psi(x)}{d\tau} &= \sum_i \frac{\partial \psi(x)}{\partial \theta_i}\frac{\partial\theta_i}{\partial d\tau} \nonumber \\ 
&= -\eta \left[ \sum_{x'\in B}\Theta(x,x') \frac{\partial \cL}{\partial \psi(x')} + \Phi(x,x') \frac{\partial \cL}{\partial \psi^*(x')} \right],
\end{align}
where 
\begin{align}
\Theta(x,x') &= \nonumber\sum_i   \frac{\partial \psi(x)}{\partial \theta_i}\frac{\partial \psi(x') }{\partial \theta_i} \\
\Phi(x,x') &= \sum_i  \frac{\partial \psi(x)}{\partial \theta_i}\frac{\partial\psi^* (x') }{\partial \theta_i}.
\end{align}
$\Theta(x,x')$ is the \emph{neural tangent kernel} (NTK) \cite{NTK}. 

Since we are using a complex-valued neural network
to represent quantum wavefunctions, we also see the appearance of $\Phi(x,x')$, which we call the \emph{Hermitian neural tangent kernel}
(HNTK), since it is Hermitian, $\Phi^*(x,x') = \Phi(x',x)$. Putting the wavefunction and its conjugate on equal footing, we write
\begin{gather}
    \label{eq:mastermatrix}
    \frac{d}{d\tau}
     \begin{bmatrix} \psi(x) \\ \psi^*(x) \end{bmatrix}
     =-\eta \sum_{x'\in B}
      \begin{bmatrix}
       \Theta(x,x') &
       \Phi(x,x') \\ 
        \Phi^{*}(x,x')&
       \Theta^*(x,x')
       \end{bmatrix}
     \begin{bmatrix} \frac{\partial \cL}{\partial \psi(x')} \\ \frac{\partial \cL}{\partial \psi^*(x')}  \end{bmatrix}
    \end{gather}
and for simplicity re-express it as 
\begin{equation}
    \label{eq:master}
    \frac{d}{d\tau} \Psi(x) = -\eta\sum_{x'\in B}{\Omega(x,x')} \frac{\partial \cL}{\partial \Psi(x')},
\end{equation}
a matrix ODE where  $\Omega(x,x')$ is the block matrix in Eq.~\ref{eq:mastermatrix}. 

We call $\Omega(x,x')$ the \emph{quantum state neural tangent kernel} (QS-NTK), as it determines the gradient descent dynamics of NNQS, and more generally of complex functions. In general, it depends on parameters $\theta_i$ and the initialization of $\psi(x)$, though we will see in appropriate limits that the QS-NTK is deterministic and frozen during training.  
See also \cite{liu2021representation}, which utilizes a  quantum NTK in the context of variational quantum circuits, and appeared while we were finishing this work.

In practice, instead of representing the wavefunction as one complex output from the neural network, it is also common to have the neural network output the real and imaginary part of the wavefunction. In this case, we have the real imaginary NNQS representation $\Psi_{RI} := (\psi_1, \psi_2)$ such that 

\begin{equation}
    \label{eq:master}
    \frac{d}{d\tau} \Psi_{RI}(x) = -\eta\sum_{x'\in B}{\Omega_{RI}(x,x')} \frac{\partial \cL}{\partial \Psi_{RI}(x')}.
\end{equation}
where $\Omega_{RI}$ is the neural tangent kernel in real imaginary representation; see the Supplementary Materials.

The QS-NTK is generic and may be applied to the various NNQS learning schemes, which correspond to the choice of loss function $L$. For Variational Monte Carlo study of ground states associated to a given Hamiltonian $H$, $L = \frac{\braket{\psi}{H \psi}}{\braket{\psi}{\psi}}$. For quantum state tomography, with observables $\ket{x}\bra{x}$ in a different basis rotation, $L = -\sum_{x} \log |\braket{x}{\psi}|^2$. For quantum state supervised learning with a target wavefunction $\psi_T$, $L = ||\psi - \psi_T||^2$. 
 In general, Eq.~\ref{eq:master} is a nonlinear ODE with rich structure.
 In this work, we focus on the quantum state supervised learning setup, which yields a linear ODE. The study of other loss functions will left for future exploration.

\emph{QS-NTK for $\infty$-NNQS.} 
Let $\psi_{\theta,N}$ be a NNQS and $\Omega_N(x,x')$ the associated quantum 
neural tangent kernel. For many architectures, the infinite QS-NTK $\Omega_\infty(x,x')$ is parameter-independent at initialization.
This is established by the kernel trick, which turns $\Omega_\infty(x,x')$ 
into an expectation value over parameters via the law of large numbers. See the Supplementary Materials for a concrete example and discussion of generality, using NTK results. Utilizing this trick generally 
requires i.i.d. parameters, a property generally spoiled by training.

Fortunately, the initialization
QS-NTK plays a special role that can resolve the issue.
Consider the linearized model associated to $\Psi(x)$,
\be
\Psi_l(x) := \Psi_0(x)+ \sum_i (\theta_i-\theta_{0,i})\frac{\partial \Psi(x)}{\partial\theta_i}\bigg|_{\theta=\theta_0}
\ee
where $\theta_0$ are the parameters at initialization and $\Psi_0(x):= \Psi(x)|_{\theta=\theta_0}$ is the initialization wavefunction. The linearized model
is the \emph{truncated} first-order Taylor expansion of $\Psi(x)$ around $\theta_0$; we emphasize the model is linear in parameters, not inputs. The QS-NTK is 
\be 
\label{eq:linNTK}
\Omega_l(x,x') = \Omega(x,x')\big|_{\theta=\theta_0},
\ee
which is a crucial conceptual result. It says that the QS-NTK $\Omega_l$ associated $\Psi_l$
is the QS-NTK $\Omega$ of $\Psi(x,x')$ at initialization, which is parameter-independent.

In summary, a $\infty$-NNQS $\Psi$ with parameter-independent QS-NTK has a linearization $\Psi_l$ that evolves under gradient descent according to 
a parameter-independent, time-independent QS-NTK $\Omega_l(x,x')$, with dynamics governed by Eq.~\ref{eq:master}, but with $\Psi$ ($\Omega$) replaced by
$\Psi_l$ ($\Omega_l$). This is a remarkable simplification.

\textbf{Quantum State Supervised Learning.}---We focus on quantum state supervised learning. This technique has important applications, such as initializing states for ground state and real time simulations, as well as understanding the representation power of the neural network architecture ~\cite{Westerhout2020}. The loss function of quantum state supervised learning for a target wavefunction $\psi_T$ is the mean square loss $L =\frac{1}{|B|} \sum_{x} |\psi_T(x)-\psi(x)|^2$. 

Given a target quantum state $\psi_T$ and a batch of samples $B$, the dynamics Eq.~\ref{eq:master}
become
\begin{align}\label{eq:l2}
\frac{d}{d\tau} \Psi_l(x) = -\frac{\eta}{|B|} \sum_{x' \in B} [\Omega M](x,x') \left[\Psi_l(x')-\Psi_T(x')\right]
\end{align}
where $M=\begin{pmatrix}0 & 1 \\ 1 & 0\end{pmatrix}$, we have used $\Psi^* = M \Psi$, and $\Psi$ ($\Omega$) have been replaced by $\Psi_l$ ($\Omega_l$) in \eqref{eq:master}.

The exact solution to this linear ODE is given by 
\begin{align}
    \Psi_{l,x}(\tau) &= \mu_x(\tau) + \gamma_x(\tau)
\end{align}
where 
\begin{align}\label{eq:analytic}
    \mu_x(\tau) &= \sum_{i,j,k} \Omega_{xi}\left(\Omega^{-1}\right)_{ij}(1-e^{-\Omega M \tau})_{jk}\Psi_{T,k} \\ 
    \gamma_x(\tau) &= \Psi_x(0)-\sum_{i,j,k}\Omega_{xi}\left(\Omega^{-1}\right)_{ij}(1-e^{-\Omega M \tau})_{jk}\Psi_k(0). \nonumber
\end{align}
We use subscripts to denote input dependence, with $x$ for a test point and Latin
indices as batch indices. For instance, $\Omega_{xi}:= \Omega(x,x_i)$ for $x_i \in B$ is an $x$-dependent $|B|$-vector and $\Omega_{ij}:= \Omega(x_i,x_j)$ for $x_i,x_j \in B$ is a
$|B|\times |B|$-matrix. The initial wavefunction appears only in $\gamma_x(t)$.

\begin{figure}[t]
\centering
        \includegraphics[width=1\linewidth]{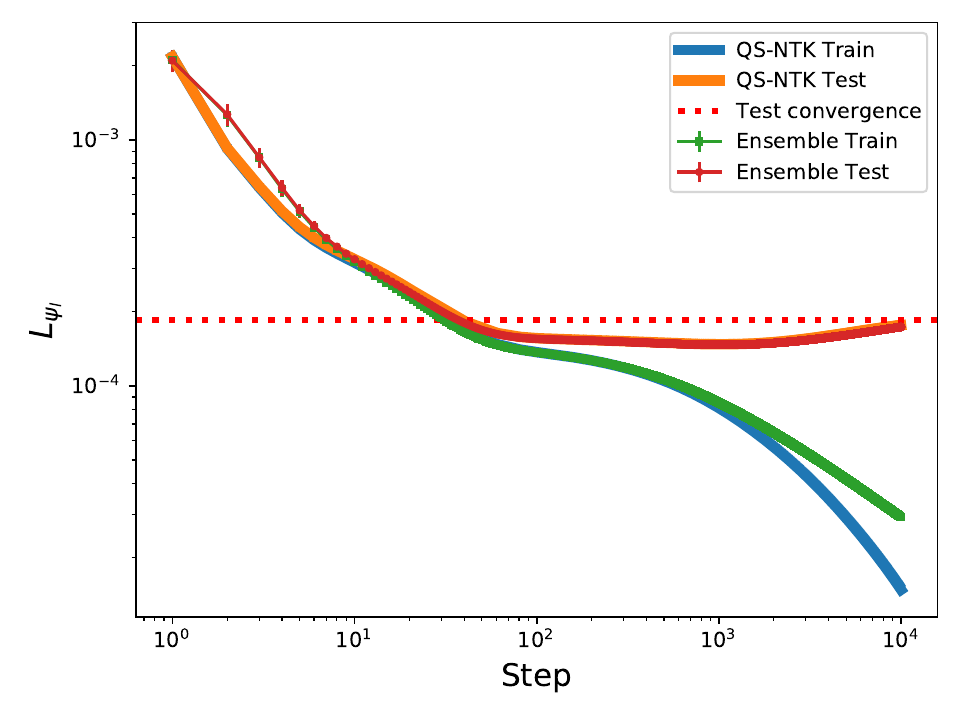}
        \includegraphics[width=1\linewidth]{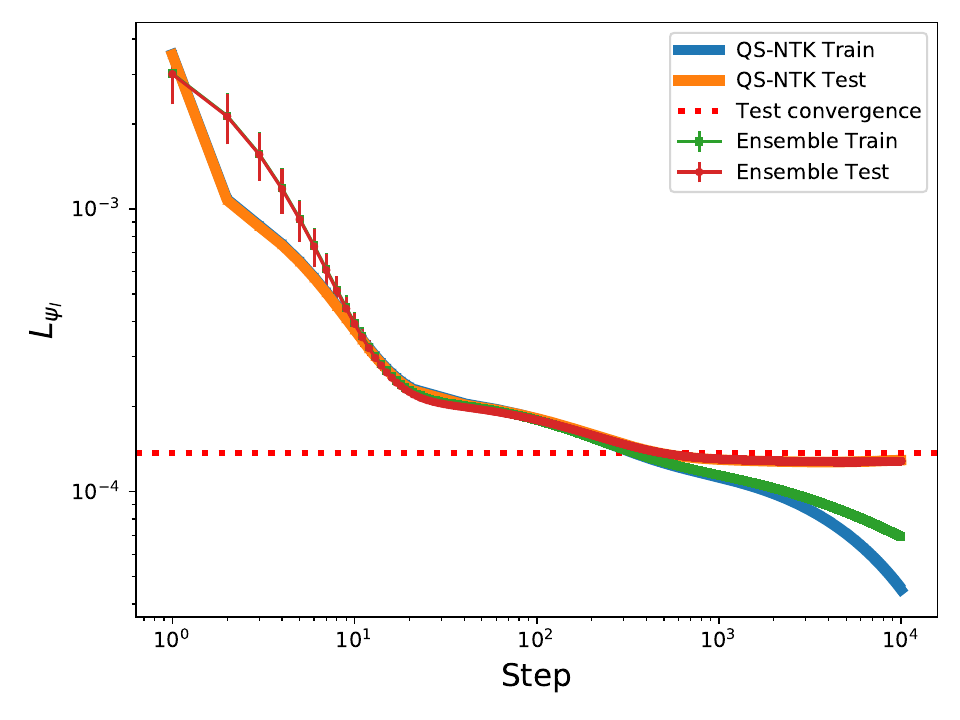}
        \caption{\label{fig:ensemble_ising} Performance of finite-width NNQS ensembles and QS-NTK predictions of $\infty$-NNQS for (a) Top: the transverse field Ising model; (b) Bottom: the Fermi-Hubbard model. The Ensemble train and test are from finite NNQS with a width equal to 5000 and ensemble size 10. The QS-NTK train and test, as well as test convergence are from infinite width NTK dynamics. The training points for both ensemble and QS-NTK cases are 2400.  }
\end{figure} 

This analytic solution for an \INNQS deserves comment.
First, when the QS-NTK is positive definite (see the Supplementary Materials), the solution converges as $\tau \to\infty$ and the converged wavefunction agrees with the target on every train point. Therefore, if the 
batch $B$ is the entire domain, the \INNQS trained with the QS-NTK perfectly reproduces the target wavefunction. This is a NNQS analog of a major result from the NTK literature, which can be understood with geometric intuition via projection from high-dimension spaces \cite{AMARI}. Equivalently, one can view $\Omega M$ as an effective Hamiltonian, in which case Eq.~\ref{eq:l2} is the analog of imaginary time evolution and converges to the ground truth.
Second, for many architectures, the expectation value of the ensemble of initial wavefunctions is 
$\bE[\Psi_x(0)]=0$, in which case $\bE[\Psi_{l,x}(\tau)] = \mu_x(\tau)$. In such a case,  $\mu_x(\tau)$
is the mean function of the ensemble at time $\tau$, and therefore $\mu_x(\infty)$ is the mean function of the infinite ensemble of converged infinite neural network quantum states.

Either $\Psi_{l,x}(\tau)$ or $\mu_x(\tau)$ could be utilized to make predictions relative to targets. This motivates two different losses,  
\begin{equation}\label{eq:mean}
   L_\mu =\frac{1}{|B|} \sum_{x' \in B} |\mu_{x'}(\infty)-\Psi_{T,x'}|^2,
\end{equation}
which uses converged mean for predictions, or 
\begin{align}
   L_{\Psi_l} =\frac{1}{K|B|}\sum_{i=1}^K \sum_{x' \in B} |\Psi^{(i)}_{l,x'}(\infty)-\Psi_{T,x'}|^2,
\end{align}
which takes the average of losses for an ensemble of $K$ linearized networks, trained to convergence, where
$\Psi^{(i)}_{l,x'}(\infty)$ is the $i^\text{th}$ network in the ensemble. Since $\bE[\gamma]=0$ as $K\to \infty$, at large $K$ we have
\begin{equation}\label{eq:variance}
L_{\Psi_l} \simeq L_\mu + \frac{1}{K|B|}\sum_{i=1}^K \sum_{x' \in B} |\gamma_{x'}^{(i)}|^2 \equiv L_\mu + L_\gamma,
\end{equation}
the last term becomes the variance of the linearized model in the $K\to \infty$ limit. Notice that Eq.~\ref{eq:analytic} shows both $L_\mu$ and $L_\gamma$ will converge both to zero on the training set in infinite time, which implies that $\infty$-NNQS will be perfectly optimized. For the test set, both $L_\mu$ and $L_\gamma$ will converge to a finite value at infinite time, which provides an indicator of the performance of the ensemble of finite neural network, in practice.

\textbf{Numerical Experiments.}---We perform numerical simulations for $\infty$-NNQS and an ensemble of finite-$N$ NNQS in two important models in quantum many-body physics, which are the spin-1/2 transverse field Ising model and the Fermi Hubbard model 
\begin{align}
    H_s &= - \sum_{\langle i,j\rangle} \sigma^z_i \sigma^z_{j} - J \sum_i \sigma^x_i,
    \label{eq:Ising} \\
H_f &= -\sum_{\langle i,j \rangle,\sigma} (c^{\dagger}_{i,\sigma} c_{j,\sigma} + h.c.) +  U\sum_i  n_{i\uparrow} n_{i\downarrow}.
\label{eq:Hubbard}
\end{align}
For the transverse field Ising model, we consider $H_s$ on a $3 \times 4$ lattice with $J=0.1$. The target state $\ket{\psi_T}$ is prepared through $\ket{\psi_T} = e^{-iH_s \tau}\ket{\psi_0}$ with $\ket{\psi_0}$ as the fully polarized state $\ket{+}^{\otimes n}$ and $\tau=2.1$. There are in total $4096$ basis elements in the target wavefunction. For the Fermi Hubbard model, we consider $H_f$ on a a $3 \times 4$ lattice with 2 spin up fermions and 2 spin down fermions. The target state in the Fermi Hubbard model is prepared through $\ket{\psi_T} = e^{-iH_f \tau}\ket{\psi_0}$, where $H_f$ has $U=8$, $\ket{\psi_0}$ is the ground state of $H_f$ with $U=4$ and $\tau=2.1$. There are $4356$ basis elements in the target wavefunction. We choose $\ket{\psi_T}$ in the above way such that they are complex-valued and related to the quench experiments with different coupling parameters in real time quantum dynamics. 

\begin{figure}[t]
        \includegraphics[width=1\linewidth]{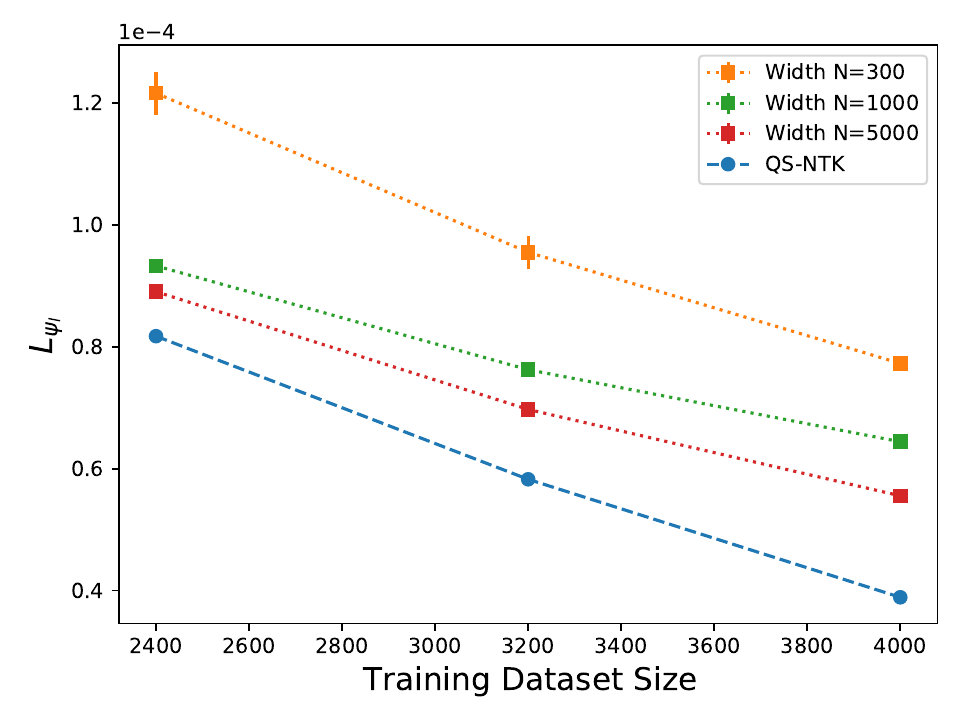}
        \includegraphics[width=1\linewidth]{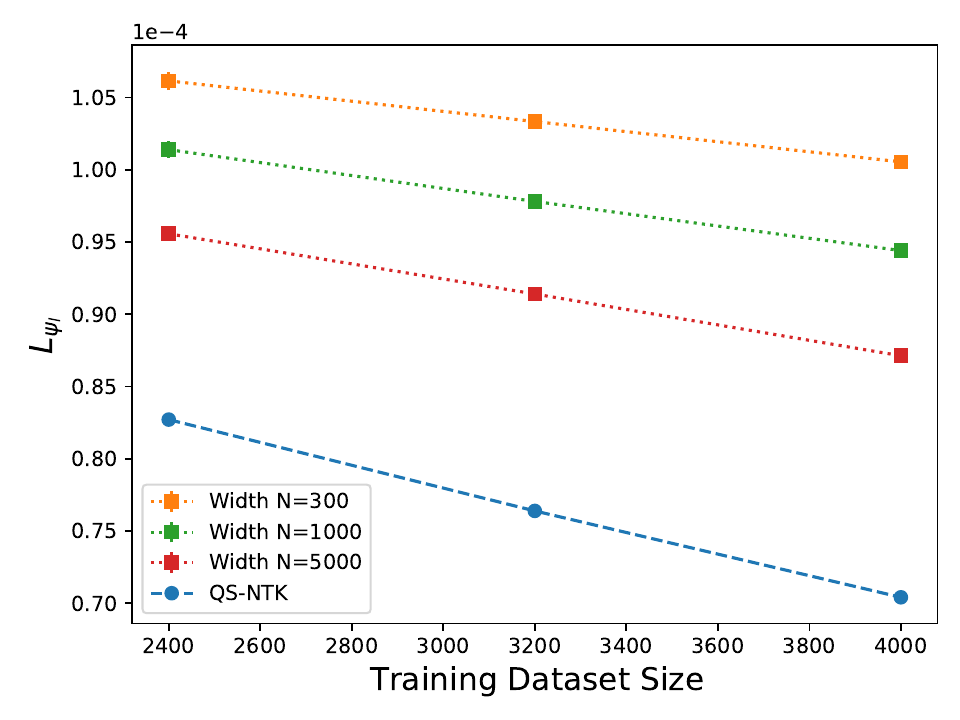}
        \caption{\label{fig:hdim_pt} Total MSE loss for various training batch sizes and finite width neural network quantum states of ensemble size 10 at training time step $\tau=10^4$ for (a) Top: the transverse field Ising model; (b) Bottom: the Fermi Hubbard model.}
\end{figure}

For the numerical simulations, we consider two independent neural networks that represent the real part and the imaginary part of the wavefunction, $\psi(x)=\psi_1(x)+i \psi_2(x)$; this is the case of decoupled dynamics discussed in the Supplementary Materials. Both $\psi_1(x)$ and $\psi_2(x)$ are single-layer fully-connected networks, i.e.,
$\frac{1}{\sqrt{N}} W_2 \sigma(\frac{W_1}{\sqrt{12}}x+b_1)+b_2$, 
with entries drawn as $W_{1,2} \sim \mathcal{N}(0,0.25)$ and $b_{1,2} \sim \mathcal{N}(0,0.01)$, $\sigma$ taken to be ReLU, and $N\in\{300,1000,5000\}$ is the dimension of the hidden layer. 

Since both models utilize $12$ lattice sites, the input is encoded in a $12$-d vector. 
For the transverse field Ising model, spin-up and spin-down configuration take values $\pm 1$. For the Fermi Hubbard model, the possibilities of a hole, spin-down, spin-up, and double occupancy take values $\in \{-1.5,-0.5,0.5,1.5\}$, respectively. For the training data set, we uniformly draw basis elements with dataset size $2400, 3200, 4000$ from the target wavefunctions, and leave the rest (the basis complement) as the test dataset. For each experiment, we train an ensemble of $10$ finite width neural network quantum states with full-batch gradient descent and compare with the quantum state neural network tangent kernel predictions. The learning rate is chosen to be $0.9$ times the maximum NTK learning rate~\cite{neuraltangents2020}, which ensure that the finite networks evolve in a linearized regime. We do not need to train the \INNQS because the exact solution Eq.~\ref{eq:analytic} makes predictions for all epochs. All simulations are implemented with \texttt{neural-tangents} library~\cite{neuraltangents2020}. 

Fig.~\ref{fig:ensemble_ising}  
compare the training dynamics of finite NNQS and \INNQS  in both the transverse field Ising model and the Fermi Hubbard model. It is shown that the finite NNQS training dynamics agree rather well with the QS-NTK predictions. The training loss for the \INNQS should drop to zero as $\tau \to \infty$, while the test losses will converge to a finite number, represented by the dashed line in the figure, which is the NTK prediction Eq. \ref{eq:variance} in the infinite time limit. Fig.~\ref{fig:hdim_pt}  
show the total MSE loss over various training dataset sizes and finite width neural network quantum state ensembles. As the training batch size increases, the overall performances of different ensembles improve as expected. As the finite width increases, the performances of the neural network quantum state ensembles converge to the NTK prediction, which is the infinite width limit.

\textbf{Conclusion.}---In this work, we introduced infinite neural network quantum states ($\infty$-NNQS). We demonstrated that ensemble average entanglement entropy bound may be computed in terms of neural network correlators. For appropriate $\infty$-NNQS, these calculations become tractable due to the NNGP correspondence. We demonstrate that certain architectures such as CosNet NNQS exhibit volume-law entanglement.
We also developed the quantum state neural tangent kernel (QS-NTK) as a general framework for understanding the gradient descent dynamics of neural network quantum states (NNQS). 
Appropriate \INNQS have parameter-independent QS-NTK at initialization, which in the linearized regime is frozen to its initialization value throughout training, leading to tractable training dynamics.
In quantum state supervised learning, we proved that training a linearized \INNQS with a positive definite QS-NTK allows for the exact recovery of any target wavefunction.
In numerical experiments, we showed that these new techniques yield accurate predictions for the training dynamics of ensembles of finite width NNQS.  Systematic studies from the infinite network literature \cite{2020arXiv200715801L} suggest that NTK or NNGP Bayesian training for \INNQS may exhibit increasing performance over finite networks.

More broadly, our work provides theoretical insights on understanding the training dynamics of neural network quantum states. It also offers practical guidance for choosing neural network architectures: convergence rates during training depend on the spectrum of the QS-NTK, evaluated on the training data. This development also opens up various interesting research directions for understanding neural network quantum states optimization in other physics contexts, such as quantum state tomography and variational Monte Carlo study of neural network quantum states. Another interesting direction is  to significantly generalize the NNQS architecture beyond the fully-connected case by using Tensor Programs~\cite{yang2021tensor}, a flexible language for connecting general architectures with NTK limits. Recently, there are applications and generalizations of neural tangent kernels to quantum computation and quantum machine learning~\cite{liu2021representation,nakaji2021quantumenhanced,shirai2021quantum,zlokapa2021quantum}, and it will be interesting to integrate QS-NTK into hybrid classical-quantum machine learning.  

\textbf{Acknowledgments.}--- We thank Ning Bao, Zhuo Chen, Bryan Clark, Adrian Feiguin, Dmitrii Kochkov, Ryan Levy, Anindita Maiti, Fabian Ruehle, Ge Yang and Tianci Zhou for discussions. J.H. is supported by NSF CAREER grant PHY-1848089. This work
is supported by the National Science Foundation under Cooperative Agreement PHY-2019786 (The
NSF AI Institute for Artificial Intelligence and Fundamental Interactions). This material is based upon work supported by the U.S. Department of Energy, Office of Science, National Quantum Information Science Research Centers, Co-design Center for Quantum Advantage (C2QA) under contract number DE-SC0012704.

\textit{Note Added}: Refs.~\cite{liu2021representation,shirai2021quantum} on quantum neural tangent kernels in the context of quantum circuits were posted to arXiv four weeks prior to this manuscript, while our work focuses on the study of neural network quantum states.

\bibliography{reference}

\clearpage
\onecolumngrid
\renewcommand\thefigure{S\arabic{figure}}  
\renewcommand\thetable{S\arabic{table}}  
\renewcommand{\theequation}{S\arabic{equation}}
\renewcommand{\thepage}{P\arabic{page}} 
\setcounter{page}{1}
\setcounter{figure}{0}  
\setcounter{table}{0}
\setcounter{equation}{0}

\def\beq{\begin{equation}}
\def\eeq{\end{equation}}

\appendix
\section{\label{app:Infinitennqs} \large{Supplementary Materials}}

\section{I. Review of Neural Tangent Kernel \label{app:NTK}}

In this Section we wish to give a brief introduction to the neural tangent kernel (NTK) \cite{NTK}, a recent breakthrough in the theoretical machine learning community
that provides new understanding of training neural networks via gradient descent. For the sake of pedagogy, we consider the case of a neural network with one-dimensional input and one-dimensional output, though the analysis trivially extends to other dimensions. We also emphasize that the notation in this section is self-contained. To that end, consider a neural network
\begin{equation}
f_\theta: \bR \to \bR
\end{equation}
with parameters $\theta$. In general $f_\theta$ is a ``big" function, in the sense that it is a composition of simpler functions. Henceforth, we suppress the subscript $\theta$ and it is to be understood that the neural network depends on parameters $\theta$. The way in which $f$ is composed out of simpler functions is known as the neural network architecture. At initialization, the parameters $\theta$ are drawn from some distribution $\theta \sim P(\theta)$ and then updated to achieve some objective, such as minimizing a scalar loss functional $\cL$.

We consider the case of training a neural network with gradient descent. In the continuous training time limit, gradient descent is given by 
\begin{equation}
\frac{df(x)}{dt} = \frac{df(x)}{d\theta_I}\frac{d\theta_I}{dt} = -\frac{df(x)}{d\theta_I}\sum_{x' \in B} \frac{dl(x')}{d\theta_I},
\end{equation}
with Einstein summation on $I$ implied.
Here $l(x')$ is a loss associated to each train point $x'$ that together sums up to $\cL$, and $B$ is a batch of train points. Note that this is full-batch
gradient descent, not stochastic gradient descent. By one more application of the chain rule, we have 
\begin{equation}
\frac{df(x)}{dt} = \frac{df(x)}{d\theta_I}\frac{d\theta_I}{dt} = -\sum_{x' \in B} \Theta(x,x') \frac{dl(x')}{df(x')}
\end{equation}
where 
\begin{equation}
\Theta(x,x') = \frac{df(x)}{d\theta_I} \frac{df(x')}{d\theta_I}
\end{equation}
is a fundamental object appearing in the gradient descent dynamics, the NTK. Due to the sum over parameters, and the fact that modern neural networks have millions of parameters, this is a complicated object that --- though fundamental --- is in general difficult to compute. Conceptually, this is the kernel function that encodes how the function-space gradient descent  update $dl(x')/df(x')$ at a train point $x'$ gets communicated to the test point $x$. Alternatively, one may think of it as the function that relates parameter-space and function-space gradient descent.

A central observation of \cite{NTK} is that the NTK simplifies significantly in an appropriate $N\to \infty$ limit, where $N$ is an appropriate width hyperparameter of the neural network. In that limit, the so-called frozen-NTK limit, $\Theta$ becomes a deterministic function $\bar \Theta$ that is training time independent. It is deterministic because the sum over parameters may be reinterpreted as an exceptation value $\bE_\theta[\cdot]$ by the law of large numbers. It is time-independent because wide neural networks  evolve as linear models \cite{NTK,lee2019wide}. This frozen-NTK limit substantially improves the tractability of the training dynamics, and if $l(x')$ is MSE loss, the dynamics are solvable.

We refer the reader to \cite{NTK,lee2019wide} for more details on this material; we have emphasized on the essentials. In our work we develop the theory to the case of neural network quantum states.

\section{II. Quantum State Neural Tangent Kernel in Real Formulation, Mixing Kernels, and Decoupled Dynamics\label{sec:appA}}

It is illustrative to also consider the system in real imaginary formulation,
writing the wavefunction as $\psi(x) = \psi_1(x) + i \psi_2(x)$. Defining
\be
\Psi_{RI}(x) := \begin{pmatrix}\psi_1  
    \\ \psi_2\end{pmatrix}, \qquad \Psi = \begin{pmatrix} 1 & i \\ 1 & -i\end{pmatrix} \Psi_{RI} =: R \Psi_{RI}, \qquad
    \frac{\partial \cL}{\partial \Psi_{RI}} = \begin{pmatrix} 1 & 1 \\ i & -i\end{pmatrix} \frac{\partial \cL}{\partial \Psi} =: R^T \frac{\partial \cL}{\partial \Psi}
\ee 
we wish to determine the gradient descent dynamics of the real and imaginary parts. From Eq.~\ref{eq:mastermatrix}, an 
appropriate change of variables gives
\begin{gather}
    \label{eq:mastermatrixR}
    \frac{d}{dt}
     \begin{bmatrix} \psi_1(x) \\ \psi_2(x) \end{bmatrix}
     =-\eta \sum_{x'\in B}
      \begin{bmatrix}
       \Theta_1(x,x') &
       \Theta_{12}(x,x') \\ 
       \Theta_{12}(x',x)&
       \Theta_2(x,x')
       \end{bmatrix}
     \begin{bmatrix} \frac{\partial \cL}{\partial \psi_1(x')} \\ \frac{\partial \cL}{\partial \psi_2(x')}  \end{bmatrix}
    \end{gather}
which we write compactly as 
\begin{equation}
    \label{eq:master_ri}
    \frac{d}{dt} \Psi_{RI}(x) = -\eta\sum_{x'\in B}{\Omega_{RI}(x,x')} \frac{\partial \cL}{\partial \Psi_{RI}(x')}.
\end{equation}
In $\Omega_{RI}(x,x')$, we note the appearance of the NTKs associated with $\psi_1(x)$ and $\psi_2(x)$,
\begin{align}
\Theta_1(x,x') &= \frac{\partial \psi_1(x)}{\partial \theta_i}\frac{\partial \psi_1(x') }{\partial \theta_i} \nonumber \\
\Theta_2(x,x') &= \frac{\partial \psi_2(x)}{\partial \theta_i}\frac{\partial \psi_2(x') }{\partial \theta_i}, 
\end{align}
as well as a new object that we call the \emph{mixing kernel}
\be
\Theta_{12}(x,x') = \frac{\partial \psi_1(x)}{\partial \theta_i}\frac{\partial \psi_2(x') }{\partial \theta_i}.
\ee 
The mixing kernel is not symmetric in $1$ and $2$, causing the transpose $\Theta_{12}^T:=\Theta_{12}(x',x)$ to also appear in $\Omega_{RI}$.
The NTK and HNTK of $\Psi$ are related to those of $\Psi_1$ and $\Psi_2$ as
\begin{align}
\Theta &= \Theta_1 - \Theta_2 + i\, \left(\Theta_{12} + \Theta_{12}^T \right) \\ \nonumber
\Phi &= \Theta_1 + \Theta_2 - i\, \left(\Theta_{12} - \Theta_{12}^T \right) \\
\end{align}
where all are functions of $(x,x')$, and QS-NTK $\Omega$ is related to $\Omega_{RI}$ by
\begin{equation}
\Omega = R \Omega_{RI} R^T
\end{equation}

The mixing kernel may be simplified by partitioning the set of parameters into subsets as 
\be 
\theta = \{\theta_1, \theta_2, \theta_s\},
\ee 
the parameters of only $\psi_1$, only $\psi_2$, and shared parameters, respectively. Then the mixing kernel simplifies to 
\be 
\Theta_{12}(x,x') = \frac{\partial \psi_1(x)}{\partial \theta_{s,i}}\frac{\partial \psi_2(x') }{\partial \theta_{s,i}},
\ee 
i.e., it only depends on the shared parameters $\theta_s$.

The mixing kernel affects dynamics: it mixes $\partial \cL/\partial \psi_2(x')$ into the update for $\psi_1$, and vice versa. We can achieve
decoupling of the dynamics of $\psi_1$ and $\psi_2$ under an additional assumption. The pointwise loss may be decomposed as
\be
\cL(x') = \cL_1(\psi_1(x')) + \cL_2(\psi_2(x')) + \cL_{12}(\psi_1(x'),\psi_2(x')),  
\ee 
according to how different pieces depend on $\psi_1$ and $\psi_2$. It is natural to call $\cL_{12}$ the mixing loss. Then we have 

\vspace{.2cm}
\noindent \textbf{Definition:} \emph{Decoupled Dynamics.} Let $\psi_1$ and $\psi_2$ be the real and imaginary parts of a NNQS with zero mixing kernel and mixing loss, $\Theta_{12}=\cL_{12}=0$. Then $\psi_1$ and $\psi_2$ evolve independently under gradient descent.

\vspace{.2cm}
\noindent Decoupled dynamics arises, for instance, in the case of state recovery studied in quantum state supervised learning.

\section{II. Deterministic Quantum State Neural Tangent Kernel: A Simple Example}

We now demonstrate a simple architecture with deterministic QS-NTK at $N=\infty$.
Consider a single-layer network  of width $N$
\begin{equation}
\psi(x) =  \frac{1}{\sqrt{N}} \sum_i (a_i + i b_i) \,\sigma(c_i x),
\end{equation}
defined by an element-wise nonlinearity $\sigma: \mathbb{R} \to \mathbb{R}$ and parameters $\theta = \{a,b,c\}$ with $a_i, b_i, c_i \sim \cN(0,1)$ has 
\begin{align} 
\Theta_N(x,x') &= \frac{1}{N} \, xx'\, \sum_i(a_i+ib_i)^2 \sigma'(c_i x) \sigma'(c_i x')\nonumber \\
\Phi_N(x,x') &= \frac{1}{N}\sum_i\bigg[2 \sigma(c_i x) \sigma(c_ix') \nonumber \\   &+ \, xx'\, (|a_i|^2+|b_i|^2) \sigma'(c_i x) \sigma'(c_i x')\bigg].
\end{align}
When the quantities in the sums are i.i.d., the law of large numbers in the $N\to \infty$ limit gives 
\begin{align} 
    \Theta_\infty(x,x') &= xx'\, \bE_\theta[(a_i+ib_i)^2 \sigma'(c_i x) \sigma'(c_i x')]\nonumber \\
    \Phi_\infty(x,x') &= 2 \bE_\theta[\sigma(c_i x) \sigma(c_ix')] \nonumber \\   &+ \, xx'\, \bE_\theta[(|a_i|^2+ |b_i|^2) \sigma'(c_i x) \sigma'(c_i x')],
\end{align}
with no sum on $i$.
In such a case $\Theta_\infty$ and $\Phi_\infty$ are parameter-independent, and therefore so is the QS-NTK $\Omega_\infty$.
While the i.i.d. criterion is not always satisfied, it usually is at initialization. This property holds for the NTK $\Theta$ for many architectures,
but a similar analysis applies also to the HNTK $\Phi$, and therefore the QS-NTK $\Omega$ should also be parameter-independent at initialization for many architectures.

More generally, obtaining a deterministic QS-NTK is simple in the decoupled limit. There an \INNQS is $\psi=\psi_1+i \psi_2$, and if the NTKs $\Theta_1$ and $\Theta_2$ of $\psi_1$ and $\psi_2$ are deterministic, then so is the QS-NTK. One simply chooses $\psi_1$ and $\psi_2$ to have architectures that realize a deterministic NTK in the infinite limit; see \cite{yangTP2} for deterministic NTKs in a wide variety of architectures. We expect that deterministic QS-NTKs are similarly general away from the decoupling limit, where one must additionally show that the mixing kernel becomes deterministic in the infinite width limit.

\section{III. Quantum State Neural Tangent Kernel Spectra for Decoupled Dynamics}

We have demonstrated that \INNQS training via gradient descent is governed by a quantum state neural tangent kernel (QS-NTK), and that convergence is related to the spectrum of the QS-NTK. We now study the spectrum of the QS-NTK in the decoupled limit, which is sufficient for ensuring the existence of positive definite QS-NTK.
In the decoupled limit,
the real and imaginary parts of the wavefunction, $\psi_1(x)$ and $\psi_2(x)$,
evolve independently under gradient descent according to their associated NTKs
$\Theta_1$ and $\Theta_2$. In this case, the QS-NTK is positive definite if $\Theta_1$ and $\Theta_2$ are positive definite.

Accordingly, we now review cases in which the limiting NTK is positive definite (PD); i.e.,
\begin{equation}
\int \int d^dx \, d^dy \, f(x) \Theta(x,y) f(y) > 0
\label{eqn:PD}
\end{equation}
for $f$ a real-valued function. Any kernel satisfying this constraint, when evaluated on a finite set of inputs, becomes a (Gram) matrix that is positive definite. Since in practice neural networks are trained on finite data sets, this Gram matrix is of particular importance, and is positive definite  when the number of parameters $\theta_i$ in the neural network is more than the number of inputs $x$. The results below are in regards to ~Eq. \ref{eqn:PD}.

Some of our results depend crucially on Bochner's theorem, which takes different forms depending 
on context.  We use that of \cite{NIPS2007_013a006f,bochnerbook}: 

\vspace{.2cm}
\noindent \textbf{Theorem.} (Bochner). A continuous translation-invariant kernel $k(x,y) = k(x-y)$ on
 $\bR^d\times \bR^d$ is positive definite if and only if it is the Fourier transform of a non-negative measure.

\vspace{.2cm}
\noindent In particular, under proper scaling, it guarantees
\begin{equation}
k(x-y) = \int_{\bR^d}\, \rho(p)\,  e^{i p\cdot (x-y)} \, dp,
\end{equation}
where $\rho(p)$ is a probability density on $\bR^d$. Therefore if the normalized Fourier transform of $k(x-y)$ is a proper probability density, $k(x-y)$ is positive definite.

We now present numerous techniques of obtaining a positive definite NTK.

\textbf{Case 1: NTK from NNGP.} Let $g:\cD\to \bR^N$ be a randomly initialized neural network, and throughout $x,x'\in \cD$. Define $f:\cD \to \bR^D$ as 
\be 
f_i(x) = \frac{1}{\sqrt{N}}W_{ij} g_j(x),
\ee 
where $W_{ij}$ is a $D\times N$ weight matrix and $W_{ij}\sim \cN(0,1)$. Colloquially, $f$ is obtained by appending a bias-less linear layer to any $N$-dimensional neural network.
Assuming each entry $g_j(x)\sim P_g$ is independent and an appropriate moment condition is satisfied, $f_i(x)$ converges to a zero-mean Gaussian process (GP) as $N\to \infty$ with
kernel
\be
K_{ij}^{\text{NNGP}}(x,x'):= \bE[f_i(x)f_j(x')] = \delta_{ij}\,\bE[g_l(x) g_l(x')],
\ee 
with no $l$-summation on the final expectation value.
In general, the NTK $\Theta$ associated to $f$ has contributions from $W$-derivatives and derivatives with respect to parameters $\theta_g$ in $g$. If the latter are frozen (not updated during training), they do no contribute to the NTK, and we have
\begin{align}
\Theta_{ij}(x,x')&=\frac{1}{N} \delta_{ij} \sum_l g_l(x) g_l(x') \\
&\stackrel{\tiny N\to \infty}{=} \delta_{ij} \,\bE[g_l(x) g_l(x')].
\end{align}
We see that 
\be 
\Theta^\infty_{ij}(x,x') = K^{\rm NNGP}_{ij}(x,x'),
\ee 
that is, when $g$ has its parameters frozen, the deterministic NTK in the infinite width limit is given by the NNGP kernel; if $W_{ij}\sim \cN(0,\sigma^2)$, they differ by a factor of $\sigma^2$.

Armed with this result, we consider a single-layer network
called Gauss-Net, defined to be $f_i(x)$ as above with $\cD = \bR^d$ and
\begin{equation}
g_j(x) = \frac{{\rm exp}(W^0_{jk} x_k)}{\sqrt{{\rm exp}(2 \sigma^2 x\cdot x/ d)}},
\end{equation}
$W^{0}_{jk} \sim \cN(0,\sigma^2/d)$, with the exponential non-linearity applied elementwise. The NNGP kernel is 
\begin{equation}
K^{{\rm NNGP}}_{ij}(x,x') = \delta_{ij}\, {\rm exp}\left(-\frac12 \frac{\sigma^2}{d}|x-x'|^2\right),
\end{equation}
which is translation invariant.
Defining $\tau=x-x'$ and freezing $\theta_g$, we have 
\be 
\Theta^\infty_{ij}(\tau) = \delta_{ij} \, \frac{\sigma}{\sqrt{d}}\int \frac{d^dp}{(2\pi)^d} e^{-\frac12 \frac{d}{\sigma^2} p\cdot p} e^{-ip\cdot \tau}.
\ee 
Since up to normalization the Fourier transform is a nice probability density (a multivariate Gaussian), Bochner's theorem guarantees that $\Theta_{ij}^\infty(\tau)$ is positive definite.

More generally, any translation invariant NNGP whose kernel satisfies Bochner's theorem defines a positive definite NTK via this mechanism of freezing all weights but those of the last layer.

\textbf{Case 2: Random Fourier Features.} Instead of appending
a neural network with a linear layer, as we just did, let us instead
prepend it with random Fourier features (RFFs) \cite{NIPS2007_013a006f}. Consider 
a neural network $g:\bR^{2d}\to \bR^D$, and consider 
the RFF map $\gamma: \bR^d \to \bR^{2d}$, where the components
of the map are
\begin{align}
\gamma_{2k} &= a_k\,  \text{cos}(2\pi b_k \cdot v) \nonumber \\
\gamma_{2k+1} &= a_k\,  \text{sin}(2\pi b_k \cdot v)
\end{align}
with $a_k \in \bR, b_k \in \bR^d$, $k \in \{0,\dots,d-1\}$. The $a_k$ and $b_k$ are tunable hyperparameters
set at initialization, and  $\gamma(v)$ lives on a hypersphere.
Since $\cos(\alpha-\beta)=\cos(\alpha)\cos(\beta)+\sin(\alpha)\sin(\beta)$,
we have 
\begin{equation}
h_\gamma(v_1-v_2) := \gamma(v_1)\cdot \gamma(v_2) = \sum_{k=0}^{d-1} a_k^2 \cos(2\pi b_k \cdot (v_1-v_2)),
\end{equation}
which, notably, is translation invariant.

We now construct another neural network that uses the RFFs.
Prepending the RFF map to $g$ we arrive at 
$f:\bR^d \to \bR^D$ 
as $f(v) = g(\gamma(v))$. Let $\Theta_g(x_1,x_2)$ be the NTK associated to $g$. Inside $f$, $g$ acts only on $x$ of 
the form $x=\gamma(v)$, i.e., points on the hypersphere. Restricted to the hypersphere, the NTK can often (e.g., if $g$ in a multi-layer perceptron) be represented as a dot product kernel, $\Theta_g(x_1,x_2)=h_g(x_1\cdot x_2)$. Then the NTK of $f$ is given by \cite{tancik2020fourfeat}
\be 
\Theta_f(v_1,v_2) = h_g(\gamma(v_1)\cdot \gamma(v_2)) = h_g(h_\gamma(v_1-v_2)).
\ee 
We see that by prepending $g$ with RFFs to obtain $f$, the NTK associated to $f$ 
is translation invariant. Additionally, convergence may be optimized by tuning the $a_k$ and $b_k$, which
in turn tunes the spectrum of $\Theta_f$; \cite{tancik2020fourfeat} demonstrated this with strong success in concrete
computer vision applications.

We emphasize instead that this technique gives another angle on PD NTKs: given an architecture $g$ with even input 
dimension, this construction yields a canonical architecture $f$ with translation-invariant NTK that may be checked
for positive-definiteness by Bochner's theorem, as we did for Gauss-net.

\textbf{Case 3: Original Literature.} The NTK was defined in \cite{NTK}, which also arrived at the first PD NTK. Let 
$f$ be a deep fully-connected network with input dimension $d$ and non-polynomial Lipschitz nonlinearity $\sigma$. Then
the restriction of the NTK to the unit sphere $S^{d-1}$ is PD.

\section{IV. Details on Entanglement Entropy Calculation}

Here we provide details on the calculation of entanglement entropy. To be specific, we consider R\'enyi-2 entropy calculation for $n=2$ in Eq.~\ref{eq:renyi}. By definition, $S_2=-\text{log} \text{Tr}[\rho_{\theta A}^{2}]$ and 

\begin{align}
    \text{Tr}[\rho_{\theta A}^{2}] &\equiv \text{Tr}_A [\rho_{\theta A}^{2}] = \text{Tr}_A [(\text{Tr}_B \ket{\psit}\bra{\psit})^{2}] \\
    &= \sum_{x_A^2, x_B^1, x_B^2}\text{Tr}_A [\ket{x_B^1 \psit}\bra{\psit x_B^1} \ket{x_A^2}\bra{x_A^2}\ket{x_B^2 \psit}\bra{\psit x_B^2}] \\
    &= \sum_{x_A^1, x_A^2, x_B^1, x_B^2}[\bra{x_A^1}\ket{x_B^1 \psit}\bra{\psit x_B^1} \ket{x_A^2}\bra{x_A^2}\ket{x_B^2 \psit}\bra{\psit x_B^2}\ket{x_A^1}] \\
    & = \sum_{x^1_A, x^1_B, x^2_A, x^2_B} [ \psi_{\theta}(x^{1,1}_{AB}) \psi_{\theta}(x^{2,2}_{AB}) \psi_{\theta}^{*}(x^{2,1}_{AB}) \psi_{\theta}^{*}(x^{1,2}_{AB}) ]
\end{align}

where $x_{AB}^{i,j}:=(x_A^i,x_B^j)$. This is also known as the replica-trick~\cite{vmc_ee,autoregressive_ee}. The ensemble average $\langle S_2 \rangle \geq -\text{log} \mathbb{E}_{\theta} \text{Tr}[\rho_{\theta A}^{2}]$, where the right hand side can be computed by

\begin{align}
    \mathbb{E}_{\theta} \text{Tr}[\rho_{\theta A}^{2}] 
    &= \mathbb{E}_{\theta} \sum_{x^1_A, x^1_B, x^2_A, x^2_B} [ \psi_{\theta}(x^{1,1}_{AB}) \psi_{\theta}(x^{2,2}_{AB}) \psi_{\theta}^{*}(x^{2,1}_{AB}) \psi_{\theta}^{*}(x^{1,2}_{AB}) ] \\ 
    &= \sum_{x^1_A, x^1_B, x^2_A, x^2_B} \mathbb{E}_{\theta} [ \psi_{\theta}(x^{1,1}_{AB}) \psi_{\theta}(x^{2,2}_{AB})/ \psi_{\theta}^{*}(x^{2,1}_{AB}) \psi_{\theta}^{*}(x^{1,2}_{AB}) ] \\ 
    &=\sum_{x^1_A, x^1_B, x^2_A, x^2_B} G^{(4)}(x_{AB}^{1,1},x_{AB}^{2,2},x_{AB}^{2,1},x_{AB}^{1,2})
\end{align}
Under the GP limit, the last line becomes the following equation and it only depends on the 2-pt correlation functions.
\begin{equation}
 \sum_{x^1_A, x^1_B, x^2_A, x^2_B} G^{(2)}(x_{AB}^{1,1},x_{AB}^{2,2}) G^{(2)}(x_{AB}^{1,2},x_{AB}^{2,1}) + G^{(2)}(x_{AB}^{1,1},x_{AB}^{1,2})G^{(2)}(x_{AB}^{2,2},x_{AB}^{2,1}) + G^{(2)}(x_{AB}^{1,1},x_{AB}^{2,1})G^{(2)}(x_{AB}^{2,2},x_{AB}^{1,2})
\end{equation}

The Von Neumann entropy can be viewed as $S_1$ of the R\'enyi-n entropy by taking $n \rightarrow 1$. In practice, it can be computed as linear combination of R\'enyi-n entropy~\cite{autoregressive_ee}. Meanwhile, $S_1 \geq S_n$ for any $n \geq 1$ from the monotonicity of the R\'enyi entropy. It implies that $\langle S_1 \rangle \geq \langle S_n \rangle$ and our calculation also provides a lower bound for the Von Neumann entropy. In addition, the Von Neumann entropy is also a limiting case of the Tsallis entropy $S^T_q = \frac{1}{1-q} (\text{Tr} \rho^q - 1)$ ~\cite{tsallis1988possible} by taking $q \rightarrow 1$. For $q \geq 2$, our approach provides an exact calculation of the ensemble average of the Tsallis entropy according to Eq.~\ref{eq:renyi}.

For the CosNet simulation, we have $\psi(x)=\psi_1(x)+i \psi_2(x)$, where are $\psi_1(x)$ and $\psi_2(x)$ are the CosNet architecture in Eq.~\ref{eq:cosnet} with weights and bias drawn independently and randomly from $a_i \sim \mathcal{N}(0,\frac{\sigma_a^2}{N}), w_{ij} \sim \mathcal{N}(0,\frac{\sigma_w^2}{d}),b_j \sim \mathcal{U}[-\pi,\pi])$. We choose $\sigma_a = 10$, $\sigma_w = \sqrt{2d}$, $N=400,1000,4000$. We consider an ensemble size 100 and for each $\psi(x)$, we construct the reduced density matrix of half of the system and compute the Von Neumann entropy. For maximum entropy, it is equal to $d_A  ln2$, where $d_A=d/2$ is the subsystem size. For the Page value of entropy between two subsystem $A$ and $B$, it follows the equation~\cite{bianchi2019typical}  

\begin{equation}
    S \approx \text{log} d_A - \frac{1}{d_A d_B} \frac{d_A^2-1}{2} 
\end{equation}
where $d_B \gg 1$. In the simulation, we take $d_A=d_B=d/2$, where $d$ is the total system size.

\end{document}